\documentclass[twocolumn,showpacs,amssymb,aps, prl]{revtex4}
\usepackage{tipa}
\usepackage{graphicx}% Include figure files
\usepackage{dcolumn}% Align table columns on decimal point
\usepackage{latexsym,bm}% bold math
\usepackage{amsmath}
\usepackage{amssymb}
\usepackage{natbib}
\usepackage{epsfig}

\begin{document}
\title{Confronting theoretical results of localized and additional surface\\plasmon resonances in silver nanoparticles with EELS measurements}
\author{Guozhong~Wang}
\email{gzw@wzu.edu.cn} \affiliation{Department of Physics, Wenzhou
University, Wenzhou 325035, China}
\date{\today}

\begin{abstract}
Raza~\emph{et al.} recently observed the extraordinarily large
energy blueshifts of localized surface plasmon resonances and
additional surface plasmon resonances in silver nanoparticles
encapsulated in silicon nitride, which are not fully understood yet.
By using the quantum model consisting of two subsystems respectively
for describing the center of mass and intrinsic motions of
conduction electrons of a metallic nanosphere and a coupling
occurred between the center of mass and conduction electrons outside
the metallic nanosphere, we firstly deduced the general energy and
line broadening size-dependence of localized surface plasmon
resonances, which removes the divergent defect of usual $1/R$
size-dependence. Secondly, we proposed that the additional surface
plasmon resonance in a metallic nanosphere originates from the
transition of the first excited state to the ground state of the
center of mass subsystem with energy levels corrected by degenerate
state pairs of the system composed of the center of mass and
intrinsic motions of conduction electrons. Then, we implemented this
generation mechanism of additional surface plasmon resonances for
silver nanoparticles encapsulated in silicon nitride and the
calculated results are well consistent with experimental results.
Furthermore, we obtained a new energy expression of localized
surface plasmon resonances, with which we successfully explained the
extraordinarily large energy blueshifts of localized surface plasmon
resonances in few-nanometer silver nanoparticles encapsulated in
silicon nitride. Finally, we calculated the localized and additional
surface plasmon resonance energies of silver nanoparticles resting
on carbon films and the calculated results perfectly explain the
experimental measurements of Scholl~\emph{et al.}. Within this
quantum model, the optical properties of metallic nanoparticles are
completely determined by degenerate or nearly degenerate state pairs
of the system composed of center of mass and intrinsic motions of
conduction electrons. Our calculations also show that additional
surface plasmon resonances play almost the equal role as localized
surface plasmon resonances for metallic nanoparticles excited by
fast moving electrons.
\end{abstract}
\maketitle

\centerline{I. INTRODUCTION}
\smallskip
Apart from the bulk plasmon resonance, the conduction electrons in a
metallic nanoparticle (NP) support another more important
self-sustaining collective oscillation, which is the well-known
localized surface plasmon resonance (LSPR) and endows metallic NPs
with particular abilities, such as local electromagnetic field rapid
oscillation inside metallic NPs, colossal enhancement of local
electric fields, extreme sensitivity to dielectric environment
variations, and squeezing light beyond the diffraction limit
\cite{ukr,gra}. These specialities of LSPRs render the manipulation
of visible light waves at the nanoscales possible. Various
applications continue to flurish in many areas, such as
surface-enhanced Raman scattering \cite{pau}, improvement of light
confinement of photovoltaic devices \cite{haa}, single-molecule
detection \cite{pzi}, single-photon generation and potential
applications for the quantum information transfer \cite{ava}, and
compact laser-driven accelerators \cite{dba}. It is also found
applications in biochemistry and biomedical fields, such as
biosensing \cite{ova}, drug delivery \cite{aky}, cancer phototherapy
\cite{sla}. Recently, real-space and real-time observation of a
plasmon-induced dissociated reaction of a single dimethyl disulfide
molecule has been realized \cite{emi}.

Paralleling with the explorations of deep physics of LSPRs, novel
findings and ideas continue to emerge. Aizpurua\,\emph{et\,al.}
proposed a mechanism to actively control the optical response of
metallic NPs by applying an external dc bias across a narrow gap
\cite{cda}; Guzzinati\,\emph{et\,al.} probed the symmetry of LSPRs
with phase-shaped electron beams \cite{ggi}; Govorov\,\emph{et\,al.}
showed that the coherent ultrafast non-dissipative energy transfer
could take place between two gold NPs with an interspaced silver
island \cite{rev}. Such advances are boosted by the development of
powerful nanoscale characterization techniques. Nowadays, electron
energy-loss spectroscopy (EELS) combined with an electron
monochromated and aberration-corrected scanning transmission
electron mircroscopy is able to achieve energy resolution even down
to $9$ meV and \r{A}ngstr\"{o}m spatial-resolution in the studies of
individual plasmonic structures \cite{wji,aga,jne,mbo}.
\begin{figure*}
\includegraphics*[width=0.80\textwidth]{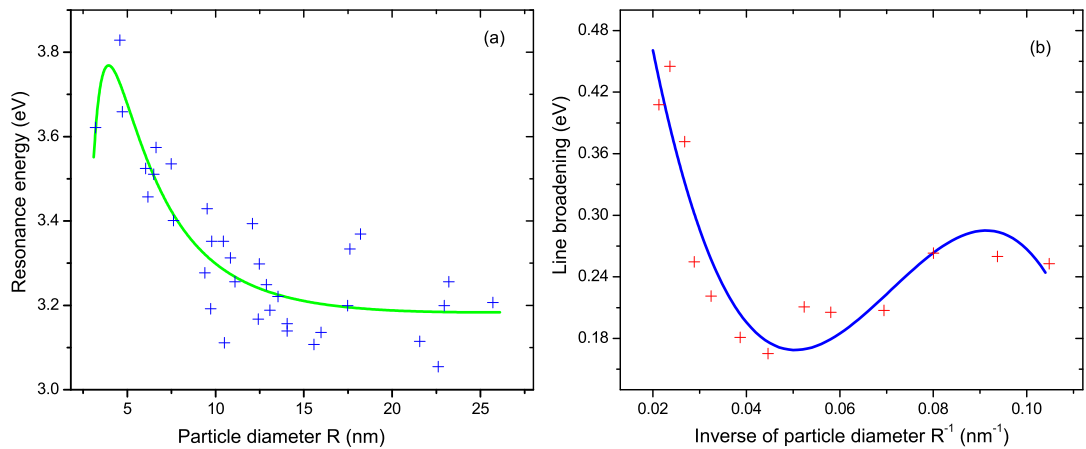}
\caption{(a)~Surface plasmon resonance energies (blue crosses)
measured in the EELS experiment for silver NPs dispersed on a
silicon nitride substrate. The fitting function is
$\hbar\Omega_q(R)=3.25-3.46/R+51.56/R^2-117.51/R^3$, which is
indicated in the green line. (b)~Measured linewidths of localized
surface plasmon resonances (red crosses) of individual silver NPs
coated with a silica shell. The fitting function is
$\hbar\gamma(R)=1.14-47.39/R+729.06/R^2-3428.79/R^3$, which is
indicated in the blue curve.} \label{fig:1} \vspace{-17pt}
\end{figure*}

The LSPRs strongly depend on size, shape, temperature, material,
dielectric environment \cite{lje,bh,kkl,avi,yoa}, which provides
multiple variables tailorable for various applications. Silver
nanostructures are ideal for plasmon studies owing to low intrinsic
losses, narrow LSPR linewidths and large optical field enhancements
\cite{wji}. For metal silver, the fully occupied $4d$ bands
described as a polarizable medium modify the frequency of the
classical Mie plasmon resonance from
$\omega_{_M}=\omega_{p}/(1+2\epsilon_{m})^{1/2}$ to
$\omega_{_M}=\omega_{p}/[Re{\,(\epsilon_{d})}+2\epsilon_{m}]^{1/2}$
\cite{anw,lj}, where $\omega_{p}=(4\pi\rho_e e^2/m)^{1/2}$ is the
plasmon frequency of bulk metal, $\epsilon_{m}$ and $\epsilon_{d}$
respectively the dielectric constant of the environment and complex
dielectric function of $4d$ bands; while $e$, $m$ and $\rho_e$ are
the electron charge, mass and density, respectively. The localized
$4d$ electrons fail to fully screen conduction electrons outside
NPs, which prevails over the spill-out effect of conduction
electrons and tends to blueshift the LSPR energies. Many different
experiments have confirmed the LSPR energy blueshifts of silver NPs
in various dielectric environments \cite{tig,kru1,ckp,mrc}. It is
obvious that atomic configurations in the vicinity of the surface
and spill-out effect of conduction electrons become more and more
prominent with the particle size decreasing, which causes the
optical properties of metallic NPs to be size-dependent.

From the classical limited mean free path effect or the calculation
of dielectric constant of silver particles by using quantum methods,
the LSPR linewidths is usually described by
$\gamma(R)=\gamma_{0}+g$\,$v_{_{F}}/R$
\cite{kru1,kru2,kaw,kwa,link,wgu1}, where $v_{_F}$ and $\gamma_0$
are Feimi velocity and intrinsic linewidth. The dimensionless $g$ is
considered to be a constant, however, quite different values of $g$
were obtained from experimental data and theoretical models
\cite{bh,ale}. The LSPR line broadening of metallic NPs is the
result of exciting single particles into electron-hole states, which
is the well-known Landau damping mechanism \cite{wgu1,jin}. By
generalized nonlocal optical response (GNOR) model,
Mortensen\,\emph{et\,al.} obtained a positive term $1/R^2$ for line
broadenings and frequency shifts \cite{mna,rso2}. Based on
experimental data \cite{ckp}, self-energy approach for particle
polarizability \cite{lia}, and dispersion relation corrected by the
spatial spreading of induced charge \cite{app}, the energy shifts of
LSPRs were found to obey the same $1/R$ size-dependence. However,
this $1/R$ law suffers from the divergent defect for few-nanometer
metallic NPs.

It has been well established that the LSPR is the dominant response
mode for metallic NPs with radii less than $20$nm to external light
excitation \cite{cth,bh,rr,kk}. By using electron energy-loss
spectroscopy (EELS), Raza~\emph{et\,al.} recently observed
additional surface plasmon resonances (ASPRs) for silver NPs
encapsulated in silicon nitride with radius range from $4$nm up to
$20$nm \cite{rso1}, which were interpreted as the combined effect of
many multipole surface plasmon resonances. While the ASPR
disappearance of silver NPs with radii below $4$nm was ascribed to
the decreasing of EELS signals from high-order modes. The measured
ASPR energies of individual silver NPs do not have a constant value,
but the multipole modes combination viewpoint of ASPRs fails to
yield a feasible energy calculation method. Furthermore, except the
dipole surface plasmon resonance, there are no other individual
multipole modes observed in experiments.

Raza~\emph{et\,al.} also observed an extraordinarily large LSPR
energy blueshift $\sim 0.9$\,eV when the particle radius decreases
from $4$\,nm to $1$\,nm. Scholl~\emph{et\,al.} observed an energy
blueshift $\sim 0.5$\,eV for individual ligand-free silver NPs
resting on carbon films \cite{sjo}, and the similar result was also
observed for silver NPs dispersed on a silicon nitride substrate
\cite{rso3}. While the negligible energy blueshifts $\sim 0.25$\,eV
appear for silver NPs embedded in solid Xe, Ar, C$_2$H$_4$
\cite{ckp,kru1}. These experimental results show that the LSPR
energy blueshifts of silver NPs strongly depend on their
surroundings. Besides the screening effect of $4d$ electrons and the
spill-out effect of conduction electrons, it is believed that the
stronger quantum confinement, single-particle excitations, nonlocal
response and numerous structural surface defects are the possible
factors jointly determining the energy blueshifts of silver NPs
\cite{cth,lje}. To understand the extraordinarily large LSPR energy
blueshifts of silver NPs is an arena to test various models or
theories. To the best of our knowledge, all theoretically predicted
energy blueshifts of silver NPs are systematically less than
experimental results \cite{rso3,mrc}.

Nowadays, the understanding of LSPR energy shifts of metallic NPs is
still poor. Quantitative predictions require the full consideration
of quantum effects, which become more and more important with the
particle size decreasing. The time-dependent density function theory
(TD-DFT) offers the possibility to address the optical response of
plasmonic systems at the quantum \emph{ab~initio} level
\cite{rer,vg}. However, the TD-DFT becomes computationally
prohibitive because their computational cost grows as fast as
$O(N_e^3)$ such that their reach is limited to systems with few
thousands of electrons. The semi-classical hydrodynamic Drude models
(HDMs) deal with differential equations of macroscopic particle
density and current density rather than single electron orbitals
gaining the advantage of numerical efficiency compared with TD-DFT,
which manifests the HDMs as a promising tool suitable to study the
optical properties of large plasmonic structures. By adding the
gradient of ground electron density to the Thomas-Fermi kinetic
energy, the hard-wall boundary condition of HDMs is removed and the
spill-out effect can be considered, which is called quantum
hydrodynamic theories (QHTs) in the literature. By assuming
electrons in different states mutually collide, a viscous stress
tensor yielding a dynamical correction to the kinetic energy
functional is expected to play the role of the Landau damping
mechanism \cite{ccr1}. It is believed that QHTs combined with
suitable electron ground-state density are able to compete with the
TD-DFT \cite{ccr2}. However, it is still challenging to build a QHT
compatible with all experimental findings \cite{yan}. It has been
shown that the plasmon resonance energies of a complex nanostructure
are equivalent to the electromagnetic interactions of plasmons from
structures with simpler configurations \cite{pre}. Therefore, a
thorough understanding of basic systems can facilitate the design of
highly sophisticated plasmonic nanostructures with desired optical
properties. In this paper, we will use a quantum model specially
constructed for metallic nanospheres to solve some unfathomed
problems associated with silver NPs.

\medskip
\centerline{II. THE QUANTUM MODEL}
\smallskip
For metallic NPs, the ionic cores can be treated as a uniform
positively charged background according to the jellium model
\cite{mat}, which is extensively adopted in numerous theoretical
models, such as TD-DFT \cite{ew}, the matrix random-phase
approximation method \cite{yc}, and field theory of quantum
plasmonics \cite{hva}. Due to strong quantum confinement, conduction
electron states are quantized into discrete levels. For a metallic
nanosphere containing $N$ atoms with radius $R$ encapsulated in the
medium with dielectric constant $\epsilon_m$, based on the jellium
model one can construct a quantum model by separating the conduction
electron coordinates into the coordinate of center of mass and the
relative coordinates (SCRM). The total Hamiltonian $\mathcal{H}_T$
of SCRM can be expressed as the sum of two sub-Hamiltonians
$\mathcal{H}_C$ and $\mathcal H_r$ respectively for describing the
collective and intrinsic motions of conduction electrons, and the
coupling $\mathcal H_c$ between center of mass and conduction
electrons outside the nanosphere, they respectively are
\cite{wg,wgu0}
\begin{eqnarray}
\mathcal{H}_C &=& \sum{}_{_{\{n\}}}(n+1/2)\hbar\Omega_p\hat{b}^{\dag}\hat{b}\nonumber\\
\mathcal{H}_r &=&
\sum{}_{_{\{\alpha\}}}\epsilon_\alpha\hat{c}^{\dag}_\alpha
\hat{c}_\alpha\nonumber\\
\mathcal{H}_c &=&
\mathcal{A}(\hat{b}^{\dag}+\hat{b})\sum{}_{_{\{\alpha,\beta\}}}d_{\alpha\beta}\hat{c}^{\dag}_\alpha
\hat{c}_\beta,
\end{eqnarray}
the sub-Hamiltonian $\mathcal{H}_C$ has the standard harmonic
oscillator structure with the frequency
$\Omega_p=\omega_{s}\sqrt{1-N_{out}/N}$, where $\omega_{s}$ is the
unique input parameter of the SCRM and slightly varies around the
classical Mie resonance frequency $\omega_{_M}$ due to numerous
surface structural defects, $N_{out}$ the number of conduction
electrons outside the nanosphere, and the coefficient $\mathcal
{A}=\frac{e^2}{4\pi\varepsilon_0R^3}\sqrt{\frac{\hbar
N}{2m\Omega_p}}$. The matrix element $d_{\alpha\beta}$ calculated
between two states $|\alpha\rangle$ and $|\beta\rangle$ of $\mathcal
{H}_{r}$ is given by
\begin{displaymath}
d_{\alpha\beta}=\langle
\alpha|\xi_z(R^3/|\vec{\,\xi}|^3-1)\Theta(|\vec{\,\xi}|-R)|\beta\rangle,
\end{displaymath}
where $\Theta(x)$ is the Heavyside step function. Within the
mean-field approximation, the energy levels and corresponding
wavefunctions of conduction electrons can be obtained by solving the
Kohn-Sham equation
\begin{equation}\label{eq3}
[-\frac{\hbar^2}{2m}\nabla^2+V_{eff}(|\vec{\,\xi}|)]\psi_{\alpha}(\vec{\,\xi})=\epsilon_{\alpha}\psi(\vec{\,\xi}),
\end{equation}
where the effective potential $V_{eff}$, usually including ionic
background potential, Hartree potential and exchange and correlation
potnetial, can be obtained by local density approximation
calculation \cite{gui2,mat}. The quantum states of total Hamiltonian
$\mathcal H_T$ can be expressed as $|I,\alpha\rangle$, where $I$ and
$\alpha$ are the quantum numbers respectively characterizing the
states of the center of mass and intrinsic motions of conduction
electrons.
\begin{figure}
\vspace{4pt}
\includegraphics*[width=0.4\textwidth]{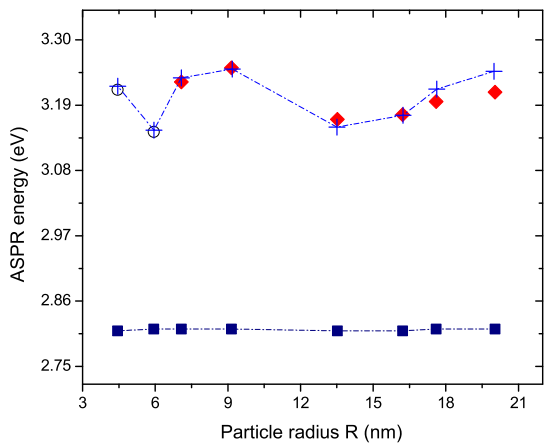}
\vspace{-6pt} \caption{The ASPR energy of silver NPs encapsulated in
homogeneous silicon nitride versus the particle radius. Red squares
and blue crosses respectively denote calculated ASPR energies and
measured ASPR energies. The first two measured ASPR energies
perfectly correspond to the calculated LSPR energies indicated by
black empty circles. The blue squares of the lower panel denote the
values of the input parameter $\hbar\omega_s$ for corresponding
silver NPs.} \label{fig:2} \vspace{-17pt}
\end{figure}
Within the SCRM framework, the LSPR frequency is given by
\begin{equation}\label{eq4}
\Omega_q(R)=\Omega_p(R)+\frac{2\mathcal{A}^2}{\hbar}\sum{}_{_{\{\alpha\beta\}}}
\mathcal{F}_{\alpha}f_{\beta}
\frac{|d_{\alpha\beta}|^2\epsilon_{\beta\alpha}}{\epsilon_{\alpha\beta}^2-(\hbar\Omega_p)^2},
\vspace{-6pt}
\end{equation}
where the sum is over all the non-degenerate state pairs
$\{|0,\alpha\rangle,|1,\beta\rangle\}$ of the total Hamiltonian
$\mathcal{H}_T$ with
$0<\epsilon_{\alpha}-\epsilon_{\beta}=\epsilon_{\alpha\beta}\ne\hbar\Omega_p$;
$f_{\beta}=1/(1+e^{(\epsilon_{\beta}-\mu)/k_{B}T})$ the Fermi-Dirac
distribution; $T$, $k_B$ and $\mu$ the electronic temperature, the
Boltzmann constant and the chemical potential respectively. We
define $\mathcal{F}_{\alpha}=1-f_{\alpha}$. The line broadening of
LSPRs caused by the Landau damping mechanism can be expressed as
\begin{equation}\label{eq5}
\hbar\gamma(R)=2\pi\mathcal
{A}^2\sum{}_{_{\{\alpha\beta\}}}\mathcal{F}_\alpha
f_\beta|d_{\alpha\beta}|^{\,2}\delta(\epsilon_{\alpha\beta}-\hbar\Omega_p),
\end{equation}
where $\delta(\epsilon_{\alpha\beta}-\hbar\Omega_p)$ is Dirac's
$\delta$ function representing the condition of energy conservation.

For a sodium nanosphere containing $1760$ atoms, the LSPR frequency
and line broadening calculated by using the SCRM are perfectly
consistent with the results of TD-DFT calculations \cite{wg}, which
shows that the SCRM is reliable to study the optical properties of
metallic NPs.

\medskip
\centerline{III. THE LSPR ENERGY SIZE-DEPENDENCE}
\smallskip
The LSPR energy and linewidth are of crucial significance in many
applications and intrinsically limit the optimization of nanosized
optical devices involving metallic NPs. Within the SCRM, the LSPR
energy shift and line broadening are determined by energy states of
conduction electrons and the input parameter $\omega_s$, which
include all effects of material, shape, size, dielectric
environment, and surface details, such as facets and vertices
\cite{mba}. The energy levels and wavefunctions of conduction
electrons can be obtained in principle by solving Eq.~(\ref{eq3}).
Thus, the effective potential of conduction electrons plays a
central role in the optical properties of metallic NPs. However, the
real effective potential of conduction electrons is so complex that
to obtain accurate energy levels and numerical wavefunctions from
Eq.~(\ref{eq3}) is extremely time-consuming and quickly becomes
prohibitive with the particle size increasing, which is the very
reason for the exploration of other calculation schemes without
resorting to single energy levels and wavefunctions.

Eqs.~(\ref{eq4}) and (\ref{eq5}) show that the size-dependences of
the LSPR energy shift and line broadening determined by
$|d_{\alpha\beta}|^2$ and the coefficient $\mathcal{A}^2$ have the
same form. Within the SCRM framework, the general size-dependence of
LSPR energies and linewidths can be worked out and the result is
\cite{s1}
\begin{equation}\label{lwf}
f(R)=f_0+\frac{A}{R}+\frac{B}{R^2}+\frac{C}{R^3}.
\end{equation}
For the situation of metallic nanospheres with different sizes
encapsulated in homogeneous medium, the coefficients $A$, $B$ and
$C$ are approximate constants with proper dimensions. The constant
$f_0$ is the intrinsic linewidth or the classical Mie plasmon
resonance energy. Besides the first two terms, two high order terms
of $1/R$ arise, which completely originate from quantum effects and
are able to eliminate the divergent defect of the usual $1/R$ law
for few-nanometer metallic NPs.

Fig.~{\ref{fig:1} shows the fits of Eq.~(\ref{lwf}) to measured
plasmon energies of silver NPs dispersed on a silicon nitride
substrate \cite{rso3}, and linewidths of individual silver NPs
coated with a silica shell \cite{bh}. The fitting curves are able to
globally describe the experimental data, especially for the
linewidths in Fig.~{\ref{fig:1}}~(b). The large energy spread of
plasmon resonances shown in Fig.~{\ref{fig:1}}~(a) is related to the
fact that part of experimental data are ASPR energies and the ASPR
and LSPR energies do not obey the same size-dependence.

The optical properties of metallic Nps are extremely sensitive to
the surface atomic configurations. The chemical control of surface
layers via ligand exchange could yield abnormal optical response
behaviors \cite{shen}, which shows that electron density tail plays
a crucial role in optical response of metallic NPs. In deducing
Eq.~(\ref{lwf}), the possible chemical bonding processes occurred
between surface atoms of metallic NPs and external molecules, which
is able to dramatically alter the spatial distribution of conduction
electron density tail, interfacial dielectric constant and the
effective potential, were not considered. Therefore, Eq.~(\ref{lwf})
can not describe the LSPR size evolution behavior of metallic NPs in
complex chemical environments.

\medskip
\centerline{IV. THE ASPRs OF METALLIC NPs}
\smallskip
The ASPR was firstly identified as a surface mode for a
semi-infinite metal by Bennett with hydrodynamic equations
\cite{bal}, and appeared in microscopic calculations for sodium
particles \cite{lia,ew}. However, the generation mechanism of ASPRs
is still unclear so far, and many quite different viewpoints of
ASPRs exist. For examples, Raza~\emph{et~al.} regard the ASPR in
silver NPs as the merger of many multipole modes; Liebsch believed
that the ASPR is the excitation that has dipolar angular character
but with an additional node in the radial distribution of the
dynamical surface screening charge compared to that of the principal
Mie plasmon oscillation \cite{lia}; while Tsuei~\emph{et~al.} deemed
the ASPR the resonance in the electron-hole pair spectrum and no
longer bearing purely dipolar character \cite{tku}. Unlike the
LSPRs, the ASPRs have neither been extensively investigated nor
attracted much attention until recently.

It is surprising that there exists a simple generation mechanism of
ASPRs within the SCRM framework. For a degenerate state pair
$|0,\alpha\rangle$ and $|1,\beta\rangle$ with
$\epsilon_{\alpha}\sim\hbar\Omega_p+\epsilon_{\beta}$, the
perturbation energy correction to states $|0,\alpha\rangle$ and
$|1,\beta\rangle$ is easy to calculate and the result is
$\pm\mathcal{A}|d_{\alpha\beta}|$ for $|0,\alpha\rangle$ and
$\mp\mathcal{A}|d_{\alpha\beta}|$ for $|1,\beta\rangle$. Because the
sub-Hamiltonian $\mathcal{H}_C$ describes collective motions of all
conduction electrons, the perturbation energy correction to each
state of a degenerate state pair is virtually the correction to
states of $\mathcal{H}_C$. It seems that these two sets of energy
corrections to the ground state $|0\rangle$ and the first excited
state $|1\rangle$ of $\mathcal{H}_C$ with opposite signs would
offset each other and produce zero results. However, the positive
energy correction $\mathcal{A}|d_{\alpha\beta}|$ to the ground state
$|0\rangle$ would increase the collective oscillation energy of
conduction electrons, which violates the principle that the ground
state of a system would have the energy as low as possible.
Therefore, the ultimate result of perturbation energy correction of
the degenerate state pair $|0,\alpha\rangle$ and $|1,\beta\rangle$
is $-\mathcal{A}|d_{\alpha\beta}|$ for the ground state $|0\rangle$
and $\mathcal{A}|d_{\alpha\beta}|$ for the first excited state
$|1\rangle$. Thus, all the degenerate state pairs would yield a new
surface plasmon resonance with the energy given by the expression
\begin{equation}\label{eq6}
\hbar\Omega_a(R)=\hbar\Omega_p(R)+2\mathcal{A}\sum{}_{_{\{\alpha\beta\}}}
\mathcal{F}_{\alpha}f_{\beta}|d_{\alpha\beta}|,
\end{equation}
where the sum is over degenerate state pair set
$\{|0,\alpha\rangle$,\,$|1,\beta\rangle\}$. It is natural to think
of this new surface plasmon resonance as the ASPR, because in
metallic NPs there only exist three kinds of plasmon resonances,
namely the LSPR, ASPR and bulk plasmon resonance.

For a metallic NP with radius $R$ less than $20$\,nm encapsulated in
homogeneous dielectric medium, the single-particle effective
potential $V_{eff}$ determining quantum states of conduction
electrons is not only complex but hard to obtain. Eq.~(\ref{eq6})
shows that the ASPR energy mainly depends on matrix elements
$d_{\alpha\beta}$, which are not sensitive to energy levels and
wavefunctions of conduction electrons. It has been shown that most
of energy levels and wavefunctions of the Schr\"odinger equation
with a Woods-Saxon-like potential are almost the same as those of
the spherical potential well of finite depth \cite{wg1}, and the
observable difference between corresponding energy levels of two
potentials focuses on energy levels well above the Fermi energy.
However, the contribution of states with high energy levels to
Eq.~(\ref{eq6}) is suppressed by the Fermi-Dirac distribution
factor. On the other hand, the wavefunctions of spherical potential
well of finite depth decay slightly faster well outside the metallic
nanosphere than those of Woods-Saxon-like potential well. However,
the deviations of matrix elements calculated by using wavefunctions
of the spherical potential well of finite depth from accurate
results are negligible for not very small nanospheres. Therefore,
the spherical potential well of finite depth equal to the sum of
Fermi energy and work function is able to substitute for the complex
single-particle effective potential to calculate the ASPR energy of
a metallic nanosphere.

There are no quantum states strictly satisfying the condition
$\epsilon_{\alpha}=\epsilon_{\beta}+\hbar\Omega_p$, thus the main
obstacle to calculate the ASPR energies by using Eq.~(\ref{eq6}) is
how to single out all degenerate state pairs from possible quantum
states $|0,\alpha\rangle$ and $|1,\beta\rangle$. The energy levels
of conduction electrons can be broadened out according to the
formula \cite{hni},
\begin{equation}\label{eq7}
\mathcal{E}(\epsilon,\epsilon_\alpha)=\frac{2}{\pi}\frac{\sqrt{\epsilon_{_T}
\epsilon_\alpha}}{(\epsilon-\epsilon_\alpha)^2+4\epsilon_{_T}\epsilon_\alpha},
\end{equation}
where $\epsilon_{_T}=(\hbar k_0)^2/2m$, and
$k_0=0.13N^{-1/3}$\r{A}$^{-1}$. Thus, the width of an energy level
changes from zero to $2\sqrt{4\epsilon_{_T}\epsilon_\alpha}$. We
define truly degenerate state pairs (TDSPs) and nearly degenerate
state pairs (NDSPs) responsible for ASPR energies and LSPR energy
shifts as state pairs with energies respectively satisfying
\begin{equation}\label{eq8}
|\hbar\Omega_p-|\epsilon_{\alpha_\beta}||\leq\mathcal{A}|d_{\alpha\beta}|,
\end{equation}
and
\begin{equation}\label{eq9}
|\hbar\Omega_p-|\epsilon_{\alpha_\beta}||
\leq(\sqrt{4\epsilon_{_T}\epsilon_\alpha}+\sqrt{4\epsilon_{_T}\epsilon_\beta}).
\end{equation}

According to the SCRM, the ASPRs originate from the degenerate state
pairs of the system. To test this viewpoint, we calculated the ASPR
energies of silver NPs encapsulated in silicon nitride under the
same settings of experiments done by Raza~\emph{et\,al} \cite{rso1}.
The conduction electron temperature is fixed at the room temperature
$T=300$K in all our calculations, and the Mie plasmon resonance
frequency is calculated by using the measured frequency-dependent
complex dielectric function of $4d$ band for bulk silver
$\epsilon_d(\omega)=(59.8+55.1i)(\omega/\omega_p)^2-(40.3+42.4i)(\omega/\omega_p)+(10.05+8.06i)$
\cite{rso3}. By taking the experimental value $3.3$ of environment
dielectric constant $\epsilon_m$ \cite{rso1}, we obtained the Mie
plasmon resonance energy $\hbar\,\omega_{_M}=2.8076$\,eV. The
calculated and measured ASPR energies are exhibited in
Fig.~\ref{fig:2}, which shows a good agreement between calculated
and experimental results. Thus, we can preliminarily conclude that
both the generation mechanism and calculation scheme of ASPRs based
on the SCRM are reasonable. For large silver NPs, Fig.~\ref{fig:2}
shows a small discrepancy between calculated and measured ASPR
energies, which is induced by using two Lorentzian or Gaussian
functions with fixed full-width at half-maximum (FWHM) $0.15$\,eV
neglecting the variation of the LSPR line broadening with the
particle size to simulate EELS spectra for determining the LSPR and
ASPR energies. The first two measured ASPR energies of silver NPs
with radii $4.44$nm and $5.94$nm are not consistent with calculated
ASPR energies. However, these two measured surface plasmon
resonances can be perfectly explained as the LSPRs.

In principle, the SCRM generation mechanism of ASPRs predicts the
existence of ASPRs for all the metallic NPs. Because there are no or
few TDSPs for few-nanometer metallic NPs, the ASPR energies normally
shift from high ASPR energies of large metallic NPs to low energies
($\sim\hbar\Omega_p(R)$) according to Eq.~(\ref{eq6}). This behavior
of ASPRs naturally explains the experimental observation that the
ASPRs of few-nanometer silver NPs encapsulated in silicon nitride
fail to be observed at high energy region containing the ASPRs of
large silver NPs. Whether or not the ASPRs of few-nanometer metallic
NPs to be experimentally observable strongly depends on the
intensity ratio of ASPRs to background signals. The ASPR energies
$\sim2.80$\,eV of silver NPs with radii $4.44$nm and $5.94$nm, and
$\sim3.11$\,eV for silver NPs with radius $3.0$\,nm were observed,
which are shown in Fig.~\ref{fig:3} and mistaken for the LSPR
energies by Raza~\emph{et al.}. These experimental results are
completely consistent with the SCRM predictions of ASPRs in silver
NPs.

However, the SCRM generation mechanism of ASPRs does not exclude the
occasional appearance of ASPRs with large resonance energies for
few-nanometer metallic NPs with special sizes. Actually, for the
silver NPs encapsulated in silicon nitride with radius
$R\sim2.62$\,nm and the input parameter $\hbar\omega_s$ ranging from
$2.795$\,eV to $2.822$\,eV, the ASPR energies vary from $3.23$eV to
$3.29$eV even larger than those of large silver NPs. Small sodium
clusters with special sizes are also able to support the ASPRs shown
by the surface loss function and dynamical polarizability calculated
respectively by using time-dependent density-functional approach and
local-density approximation method \cite{lia,ew}, which indirectly
shows that the SCRM generation mechanism of ASPRs in metallic NPs is
reasonable.

When the radius of metallic NPs increases from few nanometers to
tens of nanometers, the ASPR energy generally increases from
$\sim\hbar\Omega_p(R)$ to high ASPR energies of large metallic NPs.
When the particle radius further increases to macroscopic sizes, how
does the ASPR in metallic particles evolve? Fig.~\ref{fig:2} shows
that the ASPRs in silver NPs with macroscopic sizes seem to remain
the ASPR energies of large silver NPs. According to the SCRM
generation mechanism, the ASPR energy is determined by matrix
elements $\mathcal{A}|d_{\alpha\beta}|$ calculated between generate
state pairs, which vary with the particle radius as $\sim1/\sqrt{R}$
\cite{s1}. Therefore, the ASPR together with the LSPR in metallic
NPs would evolve into the classical Mie surface plasmon resonance
with the particle size increasing to macroscopic sizes. This
evolving mode of surface plasmon resonances is also given by the
classical electrodynamics combined with the measured dielectric
function for metallic particles excited by fast moving electrons
\cite{ftl,ale2}.

The large energy spread of surface plasmon resonances for silver NPs
resting on different substrates was observed in EELS experiments
\cite{rso3,sjo}. Besides shape variations, facets and the
interaction between particles and the substrates, it is undoubted
that the ASPRs play a significant role in the large energy spread of
surface plasmon resonances.
\begin{figure}
\vspace{4pt}
\includegraphics*[width=0.40\textwidth]{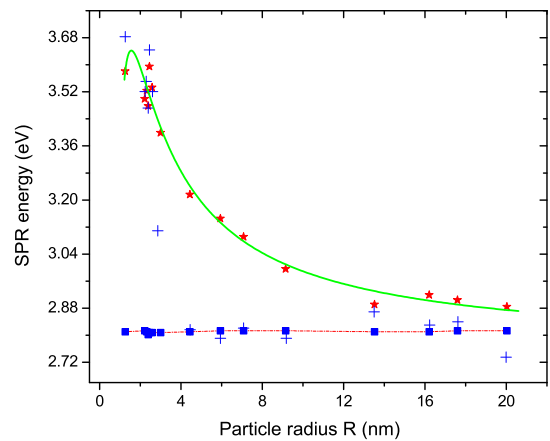}
\vspace{-6pt} \caption{The LSPR energy of silver NPs encapsulated in
silicon nitride versus the particle radius. The red pentagrams and
blue crosses respectively denote calculated and measured LSPR
energies. The green curve indicates the fitting function of
Eq.~(\ref{lwf}) to calculated LSPR energies. The blue squares of the
lower panel denote the values of the input parameter $\hbar\omega_s$
for corresponding silver NPs. In this experimental settings, the
input parameter $\hbar\omega_s$ remains almost invariable for
different silver NPs.} \label{fig:3} \vspace{-17pt}
\end{figure}

\medskip
\centerline{V. THE LSPR ENERGY SHIFTS}
\smallskip
Precise EELS experiments explicitly show that the LSPR energies of
few-nanometer silver NPs shift to higher energies by remarkable
deviations from the classical Mie surface plasmon resonance energy
\cite{sjo,rso3,rso1}. The EELS measurements also indicate that the
blueshifts of LSPR energies towards higher energies are not purely
monotonic but with a greater variety in peak locations when the
particle size decreases to few nanometers. Both relatively small
amplitude and monotonic behavior of LSPR energies predicted by known
theories and models show that the LSPR energy shift of metallic NPs
is poorly understood, and the extraordinarily large energy
blueshifts for small silver NPs have not been satisfactorily
interpreted so far \cite{mrc,sjo,rso3}.

If extremely powerful computation ability were possessed in the
future, the TD-DFT would yield LSPR energies of metallic NPs
perfectly consistent with experimental measurements. Likewise, once
the sufficiently precise energy levels and wavefunctions were
available, we also believe that the frequency expression
Eq.~(\ref{eq4}) would produce correct results of metallic NPs.
However, Eq.~(\ref{eq4}) is too sensitive to conduction electron
energy levels to produce the correct results by using energy levels
of the spherical potential well of finite depth. Fortunately,
Eq.~(\ref{eq4}) can be transformed into an alternative form by some
mathematical manipulations, which greatly lowers its sensitivity to
conduction electron energy levels. The new expression of LSPR energy
is \cite{s2}
\begin{equation}\label{eq10}
\hbar\Omega_q(R)=\hbar\Omega_p(R)\pm\frac{\pi\mathcal{A}^2}{\tau}\sum_{\{\alpha\beta\}}
(f_\beta-f_\alpha)|d_{\alpha\beta}|^2,
\end{equation}
where the signs `$+$' and `$-$' are respectively for the LSPR energy
blue and red shifts; the sum is over NDSPs with
$\epsilon_\alpha>\epsilon_\beta$. However, there are a small number
of NDSPs not contributing to the LSPR energies expressed in
Eq.~(\ref{eq10}), and such NDSPs should be excluded \cite{s3}. For
silver NPs encapsulated in silicon nitride, we calculated the LSPR
energies of silver NPs, which are measured by Raza~\emph{et al.} in
the EELS experiments. All the calculated LSPR energies and
experimental counterparts are exhibited in Fig.~\ref{fig:3}.
Nevertheless, for silver NPs with radii larger than $2.58$nm, our
calculated results are not consistent with measured ones. Because
the line broadening of LSPRs varies with the particle radius, we
think that this inconsistency is induced by the improper use of two
Gaussian functions with fixed FWHM of $0.15$eV to identify the
energies of LSPRs and ASPRs from EELS spectra \cite{rso1}. For
silver NPs with radii smaller than $2.58$nm, only LSPRs were
observed in the EELS spectra, and our calculated results are
perfectly consistent with experimental measurements naturally
explaining the extraordinarily large LSPR energy blueshifts measured
by Raza~\emph{et al.} for few-nanometer silver NPs. Furthermore, we
found that our calculated LSPR energies are well described by the
general LSPR energy size-dependence Eq.~(\ref{lwf}), which is shown
in Fig.~\ref{fig:3}.

\begin{figure*}
\vspace{4pt}
\includegraphics*[width=0.80\textwidth]{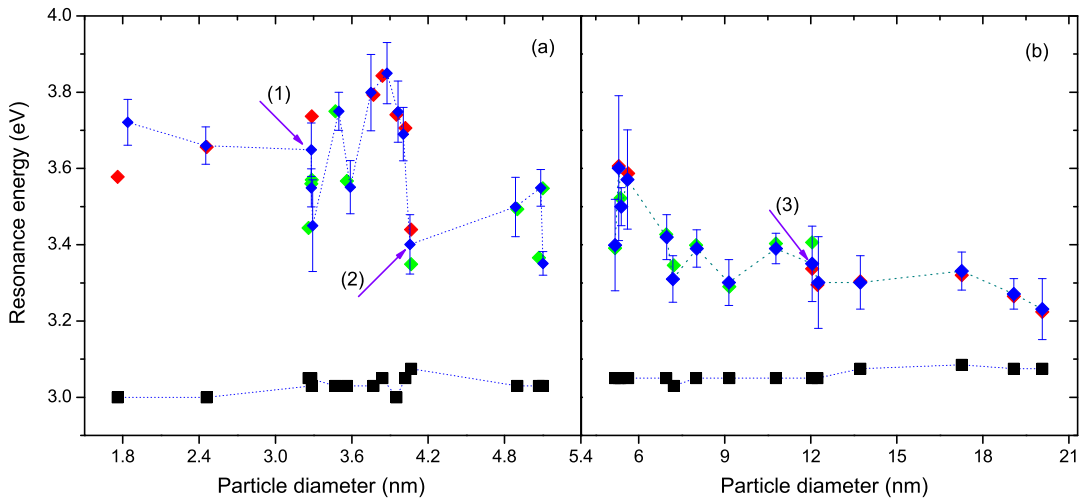}
\vspace{-6pt} \caption{The surface plasmon resonance energy versus
the particle diameter. The blue, red and green squares respectively
denote measured plasmon energies, calculated energies of LSPRs and
ASPRs. The black squares denote the values of the input parameter
$\hbar\omega_s$ for corresponding silver NPs. For clarity, all data
are depicted in two plots (a) and (b).} \label{fig:4} \vspace{-17pt}
\end{figure*}

To further test the SCRM generation mechanism of ASPRs and the LSPR
calculation formula Eq.~(\ref{eq10}), we calculated the LSPR and
ASPR energies of individual silver NPs resting on carbon films,
which were firstly studied in EELS experiments by Scholl~\emph{et
al.}. We noticed that the larger the particle size, the smaller the
influence of carbon films on the effective dielectric constant
$\epsilon_m$. Therefore, we can expect that the input parameter
$\hbar\omega_s$ for silver NPs resting on carbon films would exhibit
a larger variation amplitude than that of silver NPs encapsulated in
homogeneous silicon nitride. According to the proposed effective
dielectric constant $\epsilon_m=1.69$ \cite{sjo}, the corresponding
Mie surface plasmon resonance energy $\hbar\,\omega_{_m}=3.373$\,eV
is even larger than the measured LSPR energies of silver NPs with
diameters $\sim 20$\,nm. Considering the LSPR energy blueshifts of
silver NPs, this proposed effective dielectric constant is obviously
unreasonable. In our calculations, the input parameter varies in the
range $3.0$\,eV$<\hbar\omega_s<3.085$\,eV. The calculated energies
of LSPRs and ASPRs are shown together with measured results in
Fig.~\ref{fig:4}. Most of measured plasmon resonances (blue squares)
perfectly correspond to either calculated LSPRs (red squares) or
ASPRs (green squares). There also exists another situation. For a
silver NP with special sizes, the calculated peak locations of the
LSPR and ASPR happen to be so close to each other that they
virtually merge into one peak and are indistinguishable in EELS
experiments. We found that three measured plasmon resonances could
be perfectly explained as such merged peaks, which are indicated by
arrows in Fig.~\ref{fig:4}. We found that more than half of measured
surface plasmon resonances are ASPRs for silver NPs with the
particle diameter ranging form $3.2$\,nm to $12$\,nm. Beyond this
range, all measured plasmon resonances can be explained as LSPRs.

When a silver NP with diameter less than $2.0$\,nm, conduction
electrons have considerable possibility to stay outside the silver
NP. To substitute the spherical potential well of finite depth for
the real effective potential confining conduction electrons begins
to deteriorate, which is the reason why the first calculated LSPR
energies of Fig.~\ref{fig:3} and Fig.~\ref{fig:4} are explicitly
smaller than experimental results. Therefore, within the SCRM
framework, one has to use more precise energy levels and
wavefunctions of conduction electrons to calculate LSPR energies of
metallic NPs with radii below $1.5$\,nm.

Much larger LSPR energy blueshifts from $0.8$\,eV to $1.2$\,eV
appeared in our calculations for silver NPs resting on carbon films
with diameters in the ranges $3.26$\,nm$<D<3.32$\,nm and
$3.44$\,nm$<D<3.50$\,nm.

\medskip
\centerline{VI. SUMMARY}
\smallskip
By using the SCRM, we obtained the general size-dependence of LSPR
energies and linewidths for metallic NPs encapsulated in homogeneous
medium. Besides the LSPR and volume plasmon resonance, the ASPR is
another significant surface plasmon resonance of metallic NPs. Based
on the SCRM, we proposed that the ASPR of a metallic nanosphere
originates from all the TDSPs of the system composed of the center
of mass and intrinsic motions of conduction electrons. Then, we
implemented this SCRM generation mechanism of ASPRs in silver NPs
encapsulated in silicon nitride and explained the ASPRs measured by
Raza~\emph{et al.}, which shows that the SCRM generation mechanism
of ASPRs is reasonable. The ASPRs in metallic NPs are completely
induced by quantum effects and would evolve into the classical Mie
plasmon resonance when the particle size increases to macroscopic
sizes. The SCRM generation mechanism shows that the ASPRs in
metallic NPs almost always exist. For few-nanometer metallic NPs,
the ASPRs generally do not disappear but shift their peak locations
to low energy region $\sim\hbar\Omega_p$, which is supported by
measured ASPR energies ($\sim2.8$\,eV) of silver NPs encapsulated in
silicon nitride with radii $4.44$\,nm and $5.94$\,nm.

Within the SCRM framework, the LSPR of a metallic NP is determined
by the transition from the first excited state $|1\rangle$ to the
ground state $|0\rangle$ of the center of mass of conduction
electrons with these two energy levels corrected by all
non-degenerate states of the system. There are no essential
difference between the origins of the LSPR and ASPR in metallic NPs.
Therefore, it is somewhat strange that LSPRs can be excited by both
lights and fast moving electrons, while the ASPRs can only be
excited by fast moving electrons. To the best of our knowledge,
there are indeed no reports of ASPRs observed in experiments by
using lights to excite plasmon resonances. There are several
possible reasons for the ASPRs unobserved in the light excitation
experiments, such as relatively low experimental precision, the
light energy range being too narrow to cover LSPR and ASPR peaks, or
peak locations of the ASPR and LSPR being too close to be
distinguished. It is well known that the LSPR energies of sodium NPs
redshift, while the ASPR energies generally blueshift. Therefore, it
seems most likely to observe the ASPRs in light excitation
experiments of sodium NPs encapsulated in transparent medium.

It is usually considered that the optical properties of metallic NPs
are functions of the particle size. For silver NPs resting on a
silicon nitride substrate with the same size, the EELS measurements
show that the amplitude and linewidth of the surface plasmon
resonances can vary from particle to particle \cite{rso3}. Within
the SCRM framework, it is natural to regard the LSPR and ASPR
energies and line broadenings as the functions of the particle
radius $R$ and input parameter $\hbar\omega_s$. In a general way,
the input parameter $\hbar\omega_s$ can be expressed as
$\hbar\omega_s=\hbar\omega_{_M}[1-\delta\epsilon_m/(Re(\epsilon_d)+2\epsilon_m)]$,
where the quantity $\delta\epsilon_m$, the deviation from the
dielectric constant $\epsilon_m$ of the dielectric medium
encapsulating metallic NPs, can be further expressed as
$\delta\epsilon_m=\delta\epsilon_c+\delta\epsilon_d+\delta\epsilon_s$,
and these three terms are respectively induced by atomic
configurations in the vicinity of surfaces, shape deviation from
perfect spheres and effects related to the size of individual
metallic NPs. The first two terms in the expression of
$\delta\epsilon_m$ generally exist, but the third term
$\delta\epsilon_s$ depends on experimental settings. For metallic
NPs encapsulated in homogeneous medium, the first two terms
$\delta\epsilon_c$ and $\delta\epsilon_d$ are main effects causing
the input parameter $\hbar\omega_s$ to be different from the Mie
surface plasmon resonance energy $\hbar\omega_{_M}$, while for
metallic NPs resting on a substrate, the inhomogeneous dielectric
environments and the interactions between metallic NPs and the
substrate render the term $\delta\epsilon_s$ significant, which is
the reason why the variation amplitude of the input parameter for
silver NPs resting on carbon films is evidently larger than that of
silver NPs encapsulated in silicon nitride. It is obvious that the
input parameter $\hbar\omega_s$ has different values for different
metallic NPs even with the same size leading to different TDSP and
NDSP sets through $\hbar\Omega_p$ according to Eqs.~(\ref{eq8}) and
(\ref{eq9}), which would yield different energies and line
broadenings of LSPRs and ASPRs. The larger the variation of the
input parameter, the larger the variation of LSPR and ASPR energies
and line broadenings, which naturally explains the experimental
observations that the plasmon resonances of silver NPs resting on
carbon films have larger energy spread than that of silver NPs
encapsulated in silicon nitride. To explain the measured optical
properties of metallic NPs within the SCRM framework, besides the
sensitivity to particle sizes it is necessary to consider the
sensitivity of optical properties to the input parameter
$\hbar\omega_s$.

The primordial LSPR frequency formula Eq.~(\ref{eq4}) is extremely
sensitive to energy levels of conduction electrons. However, the
sufficiently precise energy levels of conduction electrons of a not
very small metallic NP are unavailable in most cases. Fortunately,
it is possible to transform Eq.~(\ref{eq4}) into an alternative
form, which greatly lowers the sensitivity to the conduction
electron energy levels. The calculated LSPR energies of silver NPs
encapsulated in silicon nitride are not consistent with measured
results except for silver NPs with radii below $2.58$\,nm. We think
that this inconsistency occurred for not very small silver NPs
originates from the improper employment of two Gaussian functions
with a fixed FWHM of $0.15$eV to identify the LSPR and ASPR energies
from EELS spectra. However, our calculated results perfectly explain
the large energy blueshifts of few-nanometer silver NPs encapsulated
in silicon nitride measured by Raza~\emph{et al.}.

We also calculated the LSPR and ASPR energies of silver NPs resting
on carbon films, which were studied by Scholl~\emph{et\,al.}. We
found that the ASPRs paly an important role in accounting for
experimental observations. Almost all measured surface plasmon
resonances can be explained well by the calculated LSPRs or ASPRs,
while few of measured surface plasmons correspond to the merged
resonance peaks composed of the LSPR and ASPR with similar energies
of individual silver NPs. Besides the appearance of ASPRs, the
dielectric environment of silver NPs resting on carbon films
changing with particle sizes is an important reason for large energy
spread of measured surface plasmon resonances. Our calculated
results of silver NPs being well consistent with experimental
measurements further indicate that the SCRM generation mechanism and
calculation scheme of ASPRs are quite reasonable and the LSPR energy
shifts of metallic NPs can be calculated by using the new formula
Eq.~(\ref{eq10}). We also found much larger energy blueshifts for
silver NPs resting on carbon films with particle diameters in the
ranges of $3.26$\,nm$<D<3.32$\,nm and $3.44$\,nm$<D<3.50$\,nm, which
is about twice as large as the LSPR energy blueshifts
($\sim0.5$\,eV) observed by Scholl~\emph{et al.}.

For silver NPs with radii less than $1.5$nm, our calculated LSPR
energies begin to be explicitly less than the measured ones, which
shows that the spherical potential well of finite depth is no longer
appropriate to substitute for the real effective potential
$V_{eff}$. However, the optical properties of metallic NPs with
radii less than $1.5$nm can be studied by using TD-DFT. For metallic
NPs with radii larger than $20$\,nm, the calculations based on SCRM
are no longer time-saving. However, the optical properties of large
metallic NPs can be studied by other models or theories, such as
GNOR model and various QHTs. Therefore, the SCRM could serve as a
bridge linking TD-DFT for very small metallic NPs with numerous
models or theories for large metallic NPs.

Our calculated results have shown that within the SCRM framework the
optical properties of metallic NPs are completely determined by the
TDSPs and NDSPs of conduction electrons, which play a central role
in our calculations. To the best of our knowledge, there are no
precise definitions for the NDSPs and TDSPs in the literature. We
defined them by Eqs.~(\ref{eq8}) and (\ref{eq9}), which works well
going with spherical potential well of finite depth to calculate the
optical properties of metallic NPs with sizes in the range
$1.5$\,nm$<R<20$\,nm. As a corollary of the SCRM, the LSPR energy
shift of a few-nanometer metallic NP approximately equals to its
line broadening. We hope that all theoretical predictions appeared
in this paper could be tested in the future.


\begin{thebibliography}{}
\bibitem{s1}
The radial wavefuction of a conduction electron outside the
nanosphere is
$\propto\frac{\,\,1}{\sqrt{V}}f_{\alpha}(k_{\alpha}r)e^{-k_{\alpha}r}$,
where $V=\frac{4}{3}\pi R^3$ is the volume of the metallic
nanosphere, and $k_\alpha$ the wavevector. The matrix element can be
calculated in the effective region $r>R+\delta$, where $\delta\ll R$
is the uncertainty of the particle radius, as \vspace*{-6pt}
\begin{equation}
d_{\alpha\beta}\propto\frac{1}{V}\int_{R+\delta}^{\infty}f_{\alpha}(k_{\alpha
}r)(\frac{R^3}{r^3}-1)f_{\beta}(k_{\beta}r)r^3dr.\nonumber\vspace*{-7pt}
\end{equation}
Letting $t=r/R$ and $x=t-(1+\delta/R)$, the matrix element can be
expressed as $d_{\alpha\beta}\approx{}\kappa{}R+\lambda$.
Considering the coefficient $\mathcal{A}^2\propto 1/R^3$ and
$N_{out}\propto\varrho{}R+\sigma{}R^2$ given by the similar
calculation, we obtain the size-dependence of the LSPR energy
$\hbar\Omega_q(R)=\hbar\Omega_{Mie}+\frac{A}{R}+\frac{B}{R^2}+\frac{C}{R^3}$.
The LSPR line broadening proportional to
$\mathcal{A}^2|d_{\alpha\beta}|^2$ also obeys the same relationship.
\bibitem{s2}
Firstly, we add the contribution of non-degenerate state pairs with
$\epsilon_{\alpha\beta}<0$ negligible due to the Fermi-Dirac factor
to Eq.~(\ref{eq4}), which symmetrizes the energy level indices of
Eq.~(\ref{eq4}). Secondly, by introducing a small characteristic
energy $\tau$, the summing part of Eq.~(\ref{eq4}) is transformed
into the integral \vspace{-1mm}
\begin{equation}
S\approx\frac{2\mathcal{A}^2}{\hbar\tau}\int_{-Z}^{0}[f_{\beta}-f_{\alpha}]
\frac{|d_{\alpha\beta}|^2\epsilon_{\beta\alpha}}{\epsilon_{\alpha\beta}^2-(\hbar\Omega_p)^2}
\,d\epsilon_{\alpha\beta},\nonumber \vspace{-1mm}
\end{equation}
where $Z$ equals to the sum of the work function and Fermi energy.
The integral upper limit is further extended to $Z$ and a factor 1/2
emerges. We further impose on Fermi-Dirac factor suitable properties
beyond the interval $(-Z,\,Z)$, and the integral limits can be
extended to $\pm\infty$ without causing explicit variation of the
integral value. Finally, we perform the Wick-rotation
$\epsilon_{\alpha\beta}\to i\epsilon_{\alpha\beta}$ and analytically
continue the integrand to complex plane. The obtained integral can
be calculated by using residue theorem and the result is
Eq.~(\ref{eq10}). The characteristic energy $\tau\sim
d\epsilon_{\alpha\beta}$ is taken as
$\min\{\sqrt{4\epsilon_{_T}\epsilon_\alpha}+\sqrt{4\epsilon_{_T}\epsilon_\beta}\}$
of all NDSPs.
\bibitem{s3}
To calculate the LSPR energies by using Eq.~(\ref{eq10}), we firstly
calculate the contribution of each NDSP to the LSPR energy shift by
using Eq.~(\ref{eq4}); secondly, if any two NDSPs not being TDSPs
have opposite contributions $E_{\xi\zeta}$ and $E_{\mu\nu}$
satisfying the condition
$|E_{\xi\zeta}+E_{\mu\nu}|/\max[E_{\xi\zeta},E_{\mu\nu}]<0.2$, then
such two NDSPs do not contribute to the LSPR energy shift. Finally,
the rest of NDSPs produce the LSPR energy shift according to
Eq.~(\ref{eq10}).
\end{thebibliography}
\end{document}